\documentclass[floatfix,superscriptaddress,twocolumn,aps,pre]{revtex4-2}

\usepackage{graphicx}
\usepackage{bm}
\usepackage{physics}
\usepackage[mathlines]{lineno}
\usepackage{setspace}
\usepackage[utf8]{inputenc}
\usepackage[english]{babel}
\usepackage[T1]{fontenc}
\usepackage[dvipsnames]{xcolor}
\usepackage{amssymb}
\usepackage{amsmath}
\usepackage{mathtools}
\usepackage{comment}

\usepackage{multirow}
\usepackage{array}
\newcolumntype{P}[1]{>{\centering\arraybackslash}p{#1}}
\newcolumntype{M}[1]{>{\centering\arraybackslash}m{#1}}

\newcommand*\diff{\mathop{}\!\mathrm{d}}

\begin{document}


\title{Resonant inelastic x-ray scattering in warm-dense Fe compounds beyond the SASE FEL resolution limit}

\author{Alessandro Forte}
\email{alessandro.forte@physics.ox.ac.uk}
\affiliation{Department of Physics, Clarendon Laboratory, University of Oxford, Parks Road, Oxford OX1 3PU, UK}

\author{Thomas Gawne}
\affiliation{Center for Advanced Systems Understanding (CASUS), Görlitz, D-02826, Germany}
\affiliation{Helmholtz-Zentrum Dresden-Rossendorf (HZDR), Dresden, D-01328, Germany}

\author{Karim K. Alaa El-Din}
\affiliation{Department of Physics, Clarendon Laboratory, University of Oxford, Parks Road, Oxford OX1 3PU, UK}

\author{Oliver S. Humphries}
\affiliation{European XFEL, Holzkoppel 4, 22869 Schenefeld, Germany}

\author{Thomas R. Preston}
\affiliation{European XFEL, Holzkoppel 4, 22869 Schenefeld, Germany}

\author{Céline Crépisson}
\affiliation{Department of Physics, Clarendon Laboratory, University of Oxford, Parks Road, Oxford OX1 3PU, UK}

\author{Thomas Campbell}
\affiliation{Department of Physics, Clarendon Laboratory, University of Oxford, Parks Road, Oxford OX1 3PU, UK}

\author{Pontus Svensson}
\affiliation{Department of Physics, Clarendon Laboratory, University of Oxford, Parks Road, Oxford OX1 3PU, UK}

\author{Sam Azadi}
\affiliation{Department of Physics, Clarendon Laboratory, University of Oxford, Parks Road, Oxford OX1 3PU, UK}

\author{Patrick Heighway}
\affiliation{Department of Physics, Clarendon Laboratory, University of Oxford, Parks Road, Oxford OX1 3PU, UK}

\author{Yuanfeng Shi}
\affiliation{Department of Physics, Clarendon Laboratory, University of Oxford, Parks Road, Oxford OX1 3PU, UK}

\author{David A. Chin}
\affiliation{Laboratory for Laser Energetics, University of Rochester, Rochester, NY, 14623-1299, USA}

\author{Ethan Smith}
\affiliation{Laboratory for Laser Energetics, University of Rochester, Rochester, NY, 14623-1299, USA}
\affiliation{Department of Physics and Astronomy, University of Rochester, Rochester, New York, 14627, USA}

\author{Carsten Baehtz}
\affiliation{European XFEL, Holzkoppel 4, 22869 Schenefeld, Germany}

\author{Victorien Bouffetier}
\affiliation{European XFEL, Holzkoppel 4, 22869 Schenefeld, Germany}

\author{Hauke Höppner}
\affiliation{Helmholtz-Zentrum Dresden-Rossendorf (HZDR), Dresden, D-01328, Germany}

\author{David McGonegle}
\affiliation{AWE, Aldermaston, Reading, Berkshire RG7 4PR, United Kingdom}

\author{Marion Harmand}
\affiliation{Sorbonne Université, Institut de Minéralogie, de Physique des Matériaux et de Cosmochimie, IMPMC, Museum National
d’Histoire Naturelle, UMR CNRS 7590, IRD, Paris, France}

\author{Gilbert W. Collins}
\affiliation{Laboratory for Laser Energetics, University of Rochester, Rochester, NY, 14623-1299, USA}
\affiliation{Department of Mechanical Engineering, University of Rochester, Rochester, New York, 14627, USA}
\affiliation{Department of Physics and Astronomy, University of Rochester, Rochester, New York, 14627, USA}

\author{Justin S. Wark}
\affiliation{Department of Physics, Clarendon Laboratory, University of Oxford, Parks Road, Oxford OX1 3PU, UK}

\author{Danae N. Polsin}
\affiliation{Laboratory for Laser Energetics, University of Rochester, Rochester, NY, 14623-1299, USA}
\affiliation{Department of Mechanical Engineering, University of Rochester, Rochester, New York, 14627, USA}

\author{Sam M. Vinko}
\email{sam.vinko@physics.ox.ac.uk}
\affiliation{Department of Physics, Clarendon Laboratory, University of Oxford, Parks Road, Oxford OX1 3PU, UK}
\affiliation{Central Laser Facility, STFC Rutherford Appleton Laboratory, Didcot OX11 0QX, UK}

\date{\today}

\begin{abstract}
Resonant inelastic x-ray scattering (RIXS) is a widely used spectroscopic technique, providing access to the electronic structure and dynamics of atoms, molecules, and solids. However, RIXS requires a narrow bandwidth x-ray probe to achieve high spectral resolution. The challenges in delivering an energetic monochromated beam from an x-ray free electron laser (XFEL) thus limit its use in few-shot experiments, including for the study of high energy density systems.
Here we demonstrate that by correlating the measurements of the self-amplified spontaneous emission (SASE) spectrum of an XFEL with the RIXS signal, using a dynamic kernel deconvolution with a neural surrogate, we can achieve electronic structure resolutions substantially higher than those normally afforded by the bandwidth of the incoming x-ray beam. We further show how this technique allows us to discriminate between the valence structures of Fe and Fe$_2$O$_3$, and provides access to temperature measurements as well as M-shell binding energies estimates in warm-dense Fe compounds.
\end{abstract}

\maketitle

Resonant inelastic x-ray scattering (RIXS) is a powerful spectroscopic tool that has revolutionized the field of material science. It provides detailed information on the electronic, magnetic, and orbital structures of materials with high resolution, thereby offering new opportunities for the exploration of materials and their dynamic processes including electron correlations, charge excitations, spin excitations, and phonon density of states~\cite{RevModPhys.83.705, jia2016using}. With the advent of x-ray free-electron laser (XFEL) light sources~\cite{huang2007review}, RIXS is also becoming an increasingly appealing technique for the exploration of extreme states of matter~\cite{Humphries:2020}, including of materials undergoing laser-driven dynamic compression to achieve high pressures~\cite{hicks2009laser}. 
While x-ray diffraction techniques bear witness to a rich collection of exotic physical behaviour exhibited in materials dynamically compressed to high densities~\cite{kohara2007structural}, the detailed electronic structure and excitation behaviour of such systems remain largely unexplored.

The transient nature of experimentally-realized high energy density (HED) systems necessitates the use of bright x-ray sources with short pulse durations, making XFELs particularly attractive probes. Time-resolved RIXS is appealing in this context as it can provide insight into the time evolution of the electronic structure of such systems, and access a wide range of excitation mechanisms. Traditionally, however, RIXS requires a highly monochromatic x-ray source to yield spectra with satisfactory resolution, which is at odds with the current 0.4\% bandwidth of a self-amplified spontaneous emission (SASE) XFEL~\cite{geloni2010coherence}. SASE bandwidths in the x-ray regime are typically on the order of 20-30~eV, an order of magnitude too large to resolve even basic features in the density of states. Self-seeding techniques can improve the spectral purity of the beam, but the presence of a SASE pedestal remains problematic for high-resolution RIXS measurements~\cite{liu2023cascaded}. For these reasons, monochromators are typically required, but they come with their own challenges. For one, the energy output of the x-ray pulse is drastically reduced, as most photons are discarded. In addition, the stochastic nature of the FEL generation process also leads to substantial intensity fluctuations, which further deteriorate the signal-to-noise ratio (SNR) in an integrated experiment. Importantly, in the study of transient HED systems such photon losses cannot always be recovered by longer integration times or by higher repetition rates, as the measurement is destructive, and a new system must be created for each new shot. In order to integrate multiple shots to improve overall SNR in such experiments, each shot must contain sufficient signal to be validated independently. This requirement places a fundamental constraint on the signal levels required to field a reliable RIXS diagnostic.

In this work, we present experimental results demonstrating how the electronic structure can be extracted from a RIXS measurement via a deconvolution approach that makes use of the full information contained in the stochastic SASE pulse structure of an XFEL. This idea of correlating the spectroscopic measurements with the photons source spectra has already been used in previous studies~\cite{fuller2021resonant,kayser2019core}. The resolution achieved with our approach is limited only by the resolution of the spectrometer measuring the emitted RIXS signature, and by the SNR, but not by the overall bandwidth or structure of the probe beam itself. Using a SASE pulse with bandwidth of $\sim$19~eV we show we can extract the density of states of Fe and Fe$_2$O$_3$ with resolutions of 6-9~eV. This is sufficient to observe pre-edge features in Fe$_2$O$_3$, demonstrate material specificity, and to extract the temperature of the system heated by the x-ray pulse. Our results illustrate how this correlation approach can robustly deconvolve the polychromatic signal, making the SASE spectrum of an XFEL an exploitable feature rather than an inconvenience requiring mitigation by a monochromator. Importantly, this implies that developments in XFEL technology that increase the energy of the XFEL pulse, rather than its spectral brightness at the expense of photon number, can provide a promising alternative for accessing improved, higher resolution spectroscopic data.

\section{\label{sec: RIXS as dynamic}Results}

\begin{figure}
     \centering
     \includegraphics[width=\columnwidth]{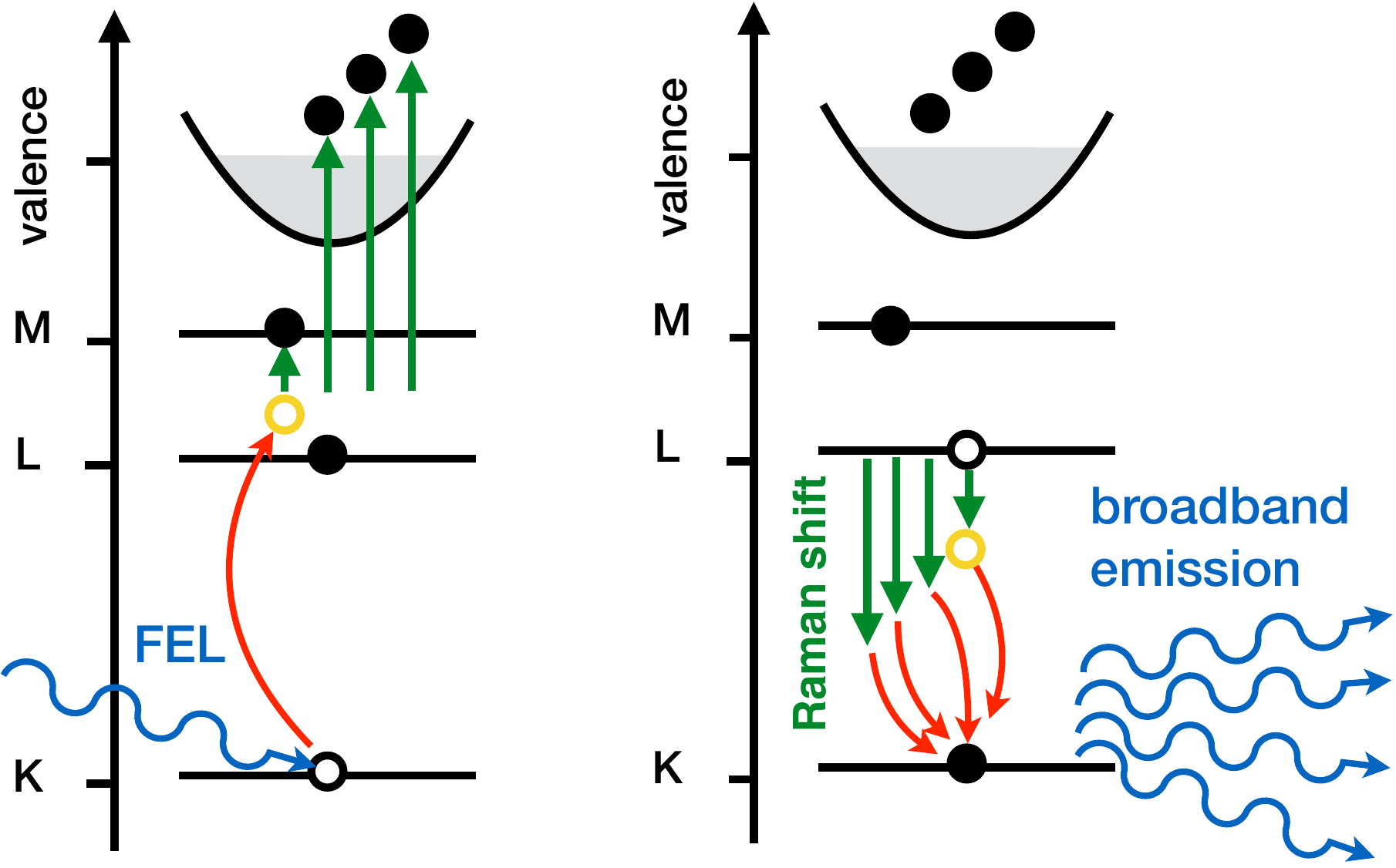}
     \caption{Schematic of the RIXS process for our cases of interest. This is composed of an energy non-conserving absorption, which excites a K-shell electron (left), followed by an energy non-conserving emission produced by the decay of a L-shell electron into the K-shell core hole (right). The overall process conserves energy. The accessible states are the vacant states in the valence band or, for finite temperature systems, in thermally ionized bound states.}
     \label{Fig:RIXS}
 \end{figure}

The RIXS cross-section for scattering into a solid angle $d\Omega$ can be written as~\cite{kramers1925streuung,kotani2001resonant}:
\begin{equation}
\begin{split}
\frac{\diff^2\sigma}{\diff \Omega \diff \omega_2} \propto 
 \sum_f & \left | \sum_n \frac{\langle f | \mathcal{D^{\prime \dagger}} | n \rangle   \langle n | \mathcal{D} | i \rangle }{E_n - \hbar\omega_1 - E_i + i\Gamma_n} \right |^2 \times \\
& \delta(E_f - E_i +  \hbar\omega_2 - \hbar\omega_1),
\label{Eq_RIXS_full}
\end{split}
\end{equation}
where $E_i$, $E_n$ and $E_f$ are the energies of the initial, intermediate and final states of the system, $\mathcal{D}$ is the transition operator~\cite{kramers1925streuung}, and $\Gamma_n$ is the lifetime of the intermediate state. The incident and outgoing photons have energies $\hbar \omega_1$ and $\hbar \omega_2$ respectively. The RIXS process is shown schematically in Fig.~\ref{Fig:RIXS}. Following the work of Humphries {\it et al.}~\cite{Humphries:2020}, we focus on the exploration of the broad structure of unoccupied valence states, rather than low-energy excitations, thus seeking to extract the density of states from the RIXS measurement.
In this case, Eq.~(\ref{Eq_RIXS_full}) can be simplified and written explicitly in terms of the density of states $\rho$, the matrix element for transitions between initial and intermediate states $M$, and the $K_{\alpha}$ line intensity $A_f$ as
\begin{multline}
\frac{\diff^2\sigma}{\diff \Omega \diff \omega_2} = \hbar \left(\frac{e}{mc}\right)^4\frac{\omega_2}{\omega_1} \sum_f A_f [1 - f_{\rm FD}(\hbar\omega_1-\hbar\omega_2+\epsilon_{L,f}; T)] \times \\
\rho(\hbar\omega_1-\hbar\omega_2+\epsilon_{L,f})
\frac{\left|M(\hbar\omega_1-\hbar\omega_2+\epsilon_{L,f})\right|^2}{(\hbar\omega_2-(\epsilon_{L,f}-\epsilon_K))^2+\Gamma_f^2},
\label{Eq:finalRIXS}
\end{multline}
where $f_{\rm FD}$ denotes the Fermi-Dirac occupation function, precluding transitions to occupied states. The energy $\epsilon_{L,f}$ denotes the binding energy of the (L-shell) electron that decays to fill the (K-shell) core hole, whose binding energy is denoted by $\epsilon_K$.
The derivation of this result can be found in the supplementary materials of Ref.~\cite{Humphries:2020}. We will use this expression as a starting point to interpret the experimental results.

\subsection{RIXS as dynamic kernel deconvolution}

\begin{figure*}
     \centering
     \includegraphics[width=0.8\textwidth]{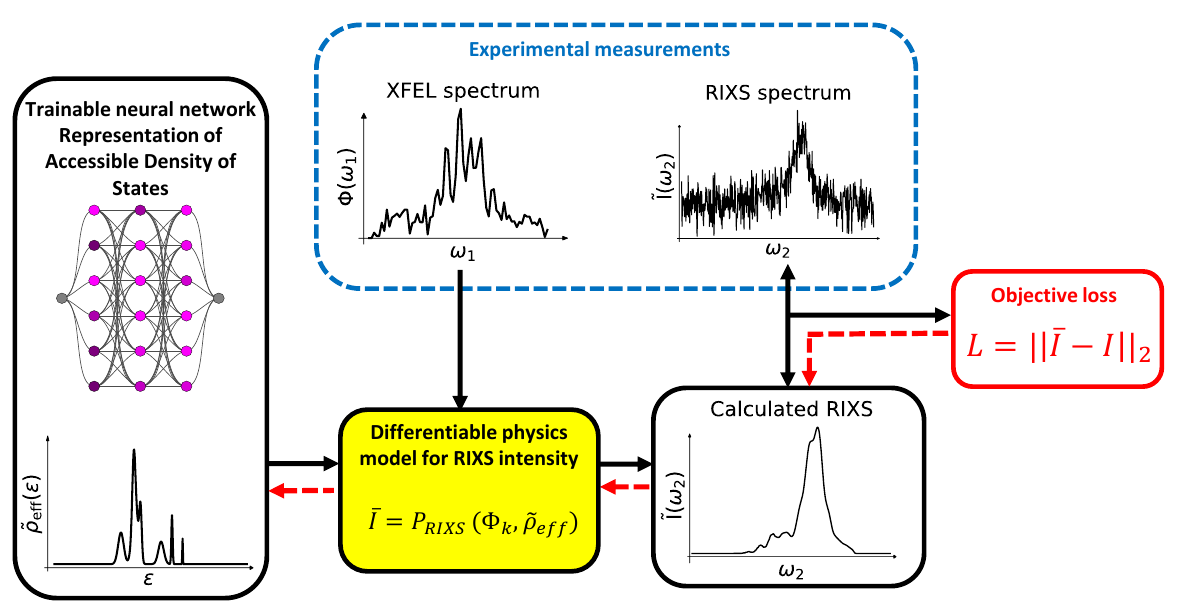}
     \caption{Representation of the scheme used to extract physical information on the electronic structure of a material from a SASE-based RIXS measurement. The resolution of the resulting density of states depends on the resolutions of the spectrometers used to measure the RIXS signal and the profile of the SASE pulse, but is independent of the SASE pulse structure or bandwidth.}
     \label{Fig:STEP}
 \end{figure*}

The intensity of the RIXS spectrum as a function of the scattered photon energy $\hbar \omega_2$, can be found by integrating the differential cross-section over all incident photon energies $\hbar \omega_1$, and over the solid angle detected for each outgoing energy $\Omega(\omega_2)$:
\begin{equation}
I(\omega_2)=\Omega(\omega_2) \int_{-\infty}^{+\infty} \hbar d\omega_1 ~\Phi(\omega_1)\partial_{\omega_2} \sigma,
\label{eq:Iomega2}
\end{equation}
with $\partial_{\omega_2} \sigma$ given by Eq.~(\ref{Eq:finalRIXS}), $\Phi(\omega_1)$ representing the incoming SASE spectrum and $\Omega(\omega_2)$ calculable from the geometry of the experimental
setup. The expression above takes the form of a sum of convolutions with a dynamic kernel $\Phi(\omega_1)$.
The kernel is dynamic because it represents the SASE pulse, which is formed of a series of narrow spikes in photon energy that change stochastically from shot to shot. If we assume a typical dataset will contain $N$ single shots indexed by $k$, with associated RIXS spectra $I_k(\omega_2)$, each will be given by Eq.~(\ref{eq:Iomega2}) using the corresponding XFEL spectra $\Phi_k(\omega_1)$.

The cross-section contains information on the vacant part of the DOS via the sum in Eq.~(\ref{Eq:finalRIXS}), including a modulation due to the energy-dependent transition matrix elements $M(\varepsilon)$. We denote this experimentally accessible quantity as the effective density of states, $\rho_{\rm eff}$, given by 
\begin{equation}
    \rho_{\rm eff}(\varepsilon) = [1-f_{FD}(\varepsilon; T)]\rho(\varepsilon)|M(\varepsilon)|^2.
    \label{Eq:DOSeff}
\end{equation}
With this notation we can describe the RIXS measurement formally as an operator $P_{\rm RIXS}$ that links the $k$ measured spectra $I_k$ to the effective DOS and the spectral FEL kernel $\Phi_k$:
\begin{equation}
    I_k(\omega_2) = P_{\rm RIXS}[\Phi_k, \rho_{\rm eff}](\omega_2).
\label{Eq:inductiveBias}
\end{equation}
The effective DOS can be found by inverting $P_{\rm RIXS}$ with respect to $\rho_{\rm eff}$:
\begin{equation}
    \rho_{\rm eff}(\varepsilon) = P^{-1}_{\rm RIXS}[I_k, \Phi_k](\varepsilon),
\end{equation}
which is valid for each $k$.

The RIXS cross-section is relatively small, and single-shot measurements typically have low SNR. This makes the use of standard inversion methods via deconvolution, such as the Richardson-Lucy method~\cite{fish1995blind}, unsuitable for single-shot analysis. Integrating over many shots ($I_k \rightarrow \bar{I}$ and $\Phi_k \rightarrow \bar{\Phi}$) can mitigate such limitations in SNR, but it also limits the resolution with which we can extract $\rho_{\rm eff}$. Alternatively, a machine learning approach could be used to construct an estimator to approximate $P_{\rm RIXS}^{-1}$ from a large labelled dataset of known pairings $(\rho_{\rm eff}, I_k, \Phi_k)$~\cite{stielow2021reconstruction}. However, given the stochastic nature of the FEL pulse profile and the complexity of the RIXS operator, collecting and validating a sufficiently large dataset of this kind can be a considerable challenge in its own right, in addition to the high complexity required for such an inversion estimator.
The lack of a robust approach to process low SNR data represents a considerable bottleneck for x-ray spectroscopy in high energy density physics applications~\cite{kayser2019core}.

\begin{figure*}[t]
     \centering
     \includegraphics[width=\textwidth]{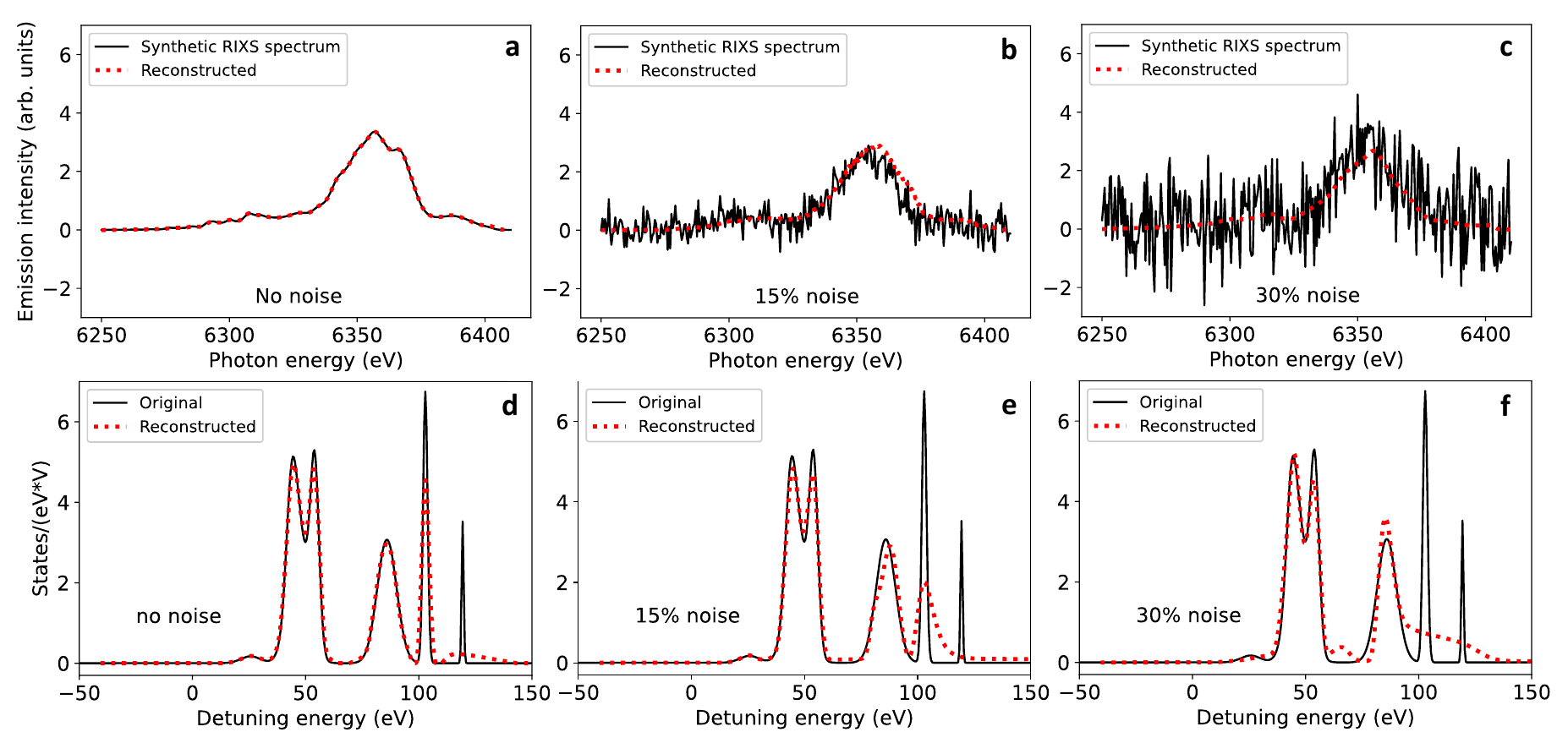}
     \caption{Reconstruction of a synthetic density of states from a RIXS spectrum at varying levels of noise, using a realistic SASE XFEL profile. A total of 50 single shots were used. The technique is robust to noise, owning to the strong inductive bias given by the model, and the stochastic nature of the SASE pulse profile $\Phi$ that allows us to oversample the DOS. The zero of the DOS energies is centred at the resonance.}
     \label{Fig:synthetic}
 \end{figure*}

Rather than searching for a general estimator to approximate the highly complex object $P_{\rm RIXS}^{-1}$, we instead look for a suitable approximation $\tilde{\rho}_{\rm eff}$ to the much simpler function $\rho_{\rm eff}$. 
The adopted procedure is illustrated in Fig.~\ref{Fig:STEP}.
We use the known forward model $P_{\rm RIXS}$ to calculate the predicted spectral intensity given a known $\Phi_k$ and the estimate $\tilde{\rho}_{\rm eff}$:
\begin{equation}
    \tilde{I}_k(\omega_2) = P_{\rm RIXS}[\Phi_k, \tilde{\rho}_{\rm eff}](\omega_2).
\end{equation}
Considering that, for our cases with low photon rates, the uncertainties of the single shots are approximately constant over $\omega_2$, the spectral intensity $\tilde{I}_k$ is then used to compute an L$_2$ loss function 
\begin{equation}
    \mathcal{L}_k = || \tilde{I}_k(\omega_2) - I_k(\omega_2)  ||_2,
\end{equation}
which provides a measure of the quality of the approximate $\tilde{\rho}_{\rm eff}$ given the observed $I_k(\omega_2)$. Improving the approximation for $\tilde{\rho}_{\rm eff}$ can now be viewed as a standard machine learning optimization problem. 
We represent $\tilde{\rho}_{\rm eff}$ using a trainable feed-forward neural network acting as a universal approximator, and implement $P_{\rm RIXS}$ in automatically differentiable form~\cite{Schoenholz2020JAXPhysics}. This allows us to use backpropagation and gradient descent to systematically improve $\tilde{\rho}_{\rm eff}$ by minimizing the objective loss $\mathcal{L}$. We explicitly use the known physics, given by Eq.~(\ref{eq:Iomega2}), to provide the inductive bias for extracting the desired electronic structure from the spectroscopic measurement.

The energy resolution with which $\tilde{\rho}_{\rm eff}$ can be found is given by the SNR of the single RIXS spectra and the energy resolutions with which $I_k$ and $\Phi_k$ are measured. While the method can, in principle, be used to optimize single shot data, we instead perform batching of the data to improve the SNR across multiple shots (see section~\ref{sec: ML approach} for details). In contrast to standard averaging approaches, the resolution is not degraded by such a merging of multiple shots, since each pairing ($I_k$, $\Phi_k$) is still considered individually in the optimization process.


\subsection{Synthetic data}



As a first validation of the method, we attempt to reconstruct synthetic DOS data using our deconvolution approach. We choose a spiky DOS, which is both challenging to tackle for traditional deconvolution algorithms, but which is also indicative of narrow bound states, $d-$band features, and resonances. 
The DOS data is fed into the forward model of Eq.~(\ref{Eq:inductiveBias}), alongside a series of realistic SASE spectra, to produce a synthetic RIXS intensity profiles. We add three levels of Gaussian noise to these spectra, with standard deviations being respectively 0\%, 15\% and 30\% of the RIXS spectra maxima. Some typical resulting spectra are shown in Figs.~\ref{Fig:synthetic}a-c. Note that at the highest level of noise, the features in the spectrum are barely recognizable.

These synthetic spectra are provided to our deconvolution scheme, alongside the corresponding SASE spectra. We show the resulting extracted DOS in Figs.~\ref{Fig:synthetic}d-f. Our approach shows good convergence of the extracted DOS for all levels of noise up to 30\%. However, the accuracy of reconstruction decreases as we raise the level of noise, with the extracted $\rho_{\rm eff}$ starting to miss small or narrow features, especially for large detuning energies. This behaviour is expected, since the area under a DOS feature and its detuning, i.e., its distance from resonance, are the two main parameters that determine the intensity of the feature in the RIXS spectrum.
Bright features closer in energy to the resonance are thus more robust to low SNR.
Another common factor that deteriorates the accuracy of the reconstructions is the presence of spurious oscillations in the extracted $\rho_{\rm eff}$. These oscillations are a typical byproduct of deconvolution, and are present on energy-scales smaller than the kernel's bandwidth. Regardless, the successful extraction of features using our approach even from spectra dominated by noise, as shown in Fig. \ref{Fig:synthetic}c, remains significant.

 \begin{figure}[t]
     \centering
     \includegraphics[width=\columnwidth]{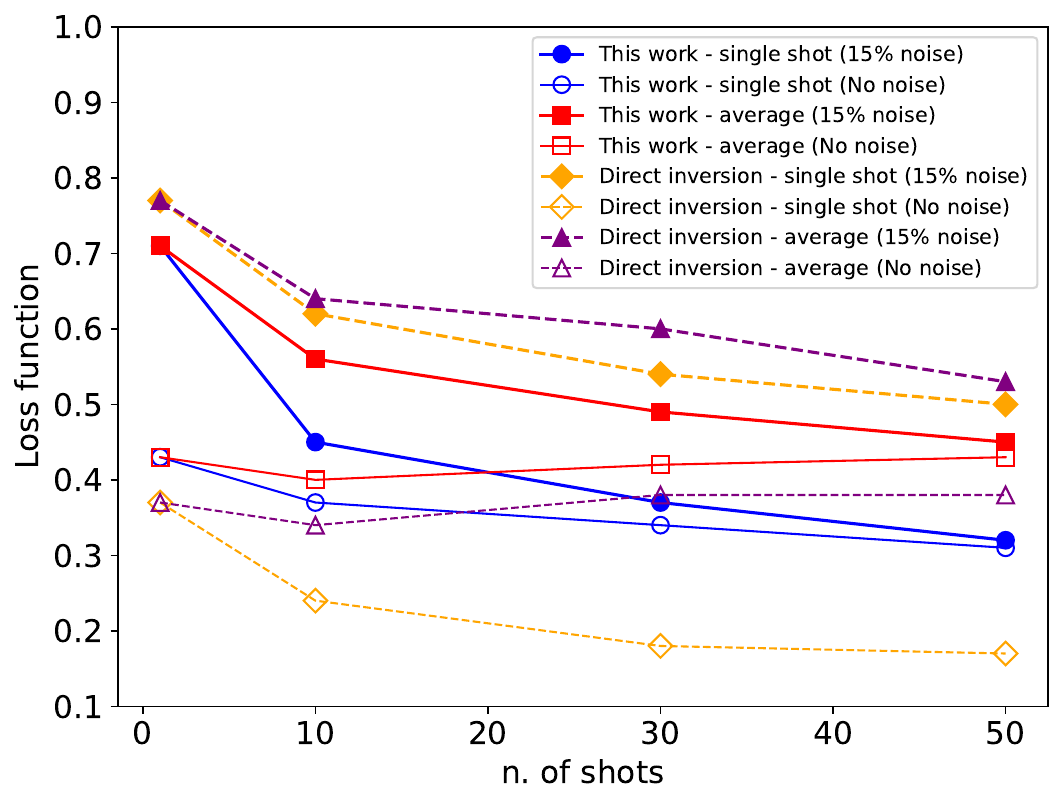}
     \caption{
     Reconstruction quality of the RIXS spectrum as a function of the number of shots N included in the inversion dataset. The DOS used is that shown in Fig.~\ref{Fig:synthetic}a. While the direct inversion works best for noiseless data, the quality rapidly deteriorates for real data with noise. In contrast, our approach is fairly insensitive to noise. The loss function employed for this plot is a weighted L$_2$ distance between the reconstructed and the original DOS, with bigger weights for the energy regions where features are present. This is achieved defining the weighted L$_2$ norm as $1/F\sum_i|d\rho_{\rm eff}/d\epsilon|(\epsilon_i)(\tilde{\rho}_{\rm eff}(\epsilon_i)-\rho_{\rm eff}(\epsilon_i))^2$, where $\rho_{\rm eff}$ is the original DOS and $F$ is a normalization factor.}
    \label{Fig:numberofshots_studies}
 \end{figure}


To demonstrate the efficacy of our method, in Fig.~\ref{Fig:numberofshots_studies} we show the quality of the $\rho_{\rm eff}$ reconstruction as a function of the number of shots in the dataset for four different approaches and two noise levels.
The blue and red curves employ the paradigm described in Fig.~\ref{Fig:STEP} to reconstruct $\rho_{\rm eff}$: with blue circles we show the results using all individual experimental pairs ($I_k$,$\Phi_k$), while the results using average values ($\bar{I}$,$\bar{\Phi}$) are shown with red squares.
We also directly invert Eq.~(\ref{Eq:inductiveBias}) via standard numerical techniques for comparison (details given in section~\ref{sec:numerical approach}), both for the individual pairs (diamond markers) and for the averaged spectra (triangle markers).
We find that while the direct inversion method achieves the best results in the absence of noise, its performance degrades rapidly as noise is added. This is due to the characteristics of $P_{{\rm RIXS}}$, which maps very different $\rho_{\rm eff}$ into similar RIXS spectra (contraction map), making its inversion an ill-conditioned problem. On the other hand, our approach shows little variation on the quality of the reconstructed DOS with noise, and is thus more robust. In particular, we observe that with our method the results obtained for a 15$\%$ noise level converge to those achieved without noise already for a relatively restrained N$\approx$50.
The comparison between single shot and averaged analysis indicates that averaged calculations suffice for broad features, but that shot-by-shot analysis is required to capture finer structure, and that the averaging process leads to a loss in resolution. This can be explained considering that deconvolution is more efficient with spiky XFEL spectra, characteristic of the single shot case, rather than with relatively broad Gaussian XFEL spectra, which occur in averaged calculations.  Specifically, in the absence of noise, where no gain on the SNR is obtained through averaging, the use of the averaged spectra deteriorates the quality of the reconstructions as N grows.



\subsection{Experimental data}

\begin{figure*}[t]
     \centering
     \includegraphics[width=\textwidth]{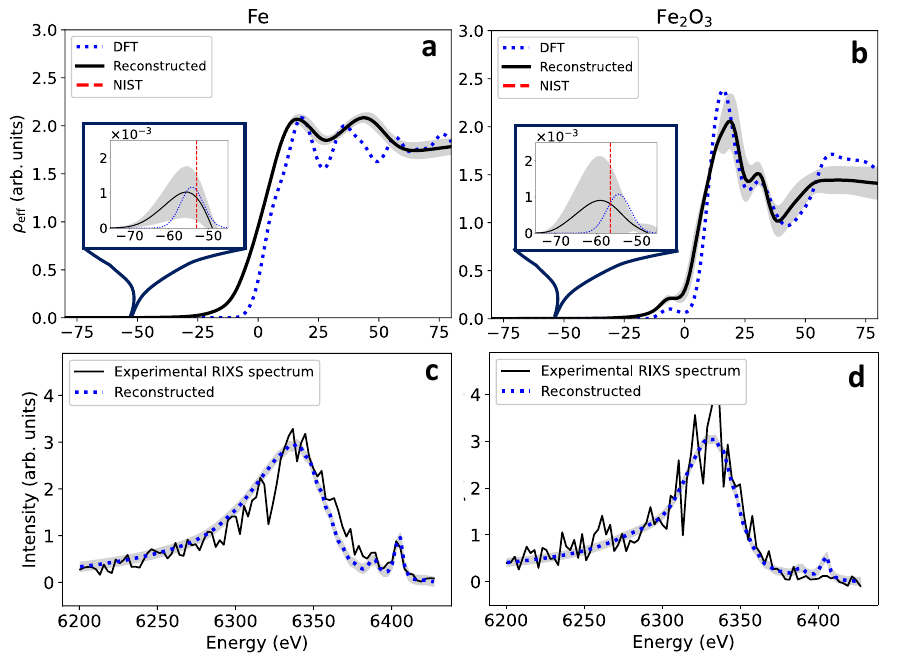}
     \caption{Measured accessible density of states of Fe and Fe$_2$O$_3$ (top row), reconstructed from the RIXS measurements, along with their respective experimental spectra (bottom row). The grey bands indicate the standard deviation. The measured data are compared with theoretical DFT calculations, for which we used a smearing width of 7 eV~\cite{parr1995density}. The insets show the reconstruction of the M-shell vacant states, located around -55~eV. This feature is only present if the M-shell is thermally ionized by the x-ray beam, thus allowing us to estimate the temperature of the sample. The red dashed vertical line indicates a reference M-shell binding energy, taken from~\cite{lemmon2010nist}.}
     \label{Fig:reconstructed}
\end{figure*}

We now turn to the analysis of experimental RIXS spectra, which have been obtained at the HED instrument of the European XFEL~\cite{zastrau2017conceptual}.
The measurement was performed using x-rays at a photon energy of 7060~eV, focused to spot sizes of 7-10~$\mu$m onto samples of Fe and Fe$_2$O$_3$. The total pulse energy in the beam ranged between 500-1000~$\mu$J, but it was constrained to be between 700-800~$\mu$J for the analysed data in order to limit the sample thermodynamic variations. The focal spot size was optimized {\it in-situ} to maximize x-ray heating, diagnosed via the observed emission from the Fe M-shell. 
The targets consisted of 20~$\mu$m thick freestanding Fe foils, and 15~$\mu $m thick Fe$_2$O$_3$, deposited on 50~$\mu$m of plastic. These thicknesses correspond to a single absorption length or below at the photon energies used, and they were chosen to maintain a uniform temperature from x-ray heating.
The FEL was operated in SASE mode~\cite{geloni2010coherence}, with an average pulse duration of 40~fs and a spectral bandwidth of around 19~eV full-width-half-maximum (FWHM).
The photon energy was tuned to lie just below the Fe K-edge to ensure that RIXS was the dominant scattering process. By ensuring that the bandwidth of the pulse is sufficiently close to the K$_{\beta}$ transition energy (1s-3p) we further ensure that the ionization of the $3p$ state can be measured. The $3p$ state is fully occupied in the ground state, and is only depopulated in the experiment due to the heating of the electrons via the intense x-ray irradiation. The RIXS signal was measured using a cylindrically bent Highly Annealed Pyrolytic Graphite (HAPG) spectrometer in the von Hamos configuration~\cite{preston2020design}, coupled to a Jungfrau detector. The experimentally determined resolution of the spectrometer and setup, including crystal resolution, pixel size effects and source size, was 5.5~eV. The spectrum of the SASE beam was determined on a shot-to-shot basis via a Si beamline spectrometer with resolution of 0.3 eV~\cite{kujala2020hard}.

Examples of single-shot experimental spectra are shown in Fig.~\ref{Fig:reconstructed}, alongside with the corresponding extracted electronic structure. We see that the typical level of noise for a single-shot RIXS spectrum is on the order of 15\%, comparable to the cases examined for the synthetic data. The experimental reconstructions of $\rho_{\rm eff}$ are compared with density functional theory calculations (see section~\ref{sec: dft calculations}), which allow us to identify various features in the experimental data, and to evaluate the resolution with which $\rho_{\rm eff}(\varepsilon)$ can be extracted in practice, given our experimental setup. The experimental $\rho_{\rm eff}$ have been reconstructed using our dynamic kernel deconvolution scheme over approximately 18,000 RIXS shots for each material. We conducted 6 independent fitting processes, for both materials, with different random seeds for the NN. The average and standard deviation of the resulting outputs have been taken, respectively, as the best estimate for $\rho_{\rm eff}$ and its error. The error on the experimental RIXS spectra due to noise, which is propagated through to $\rho_{\rm eff}$, becomes negligible because of the large amount of experimental shots. Hence, the error is due only to the stochasticity of the NN optimization.


\subsection{Estimation of the resolution}

\begin{figure}[t]
     \centering
     \includegraphics[width=\columnwidth]{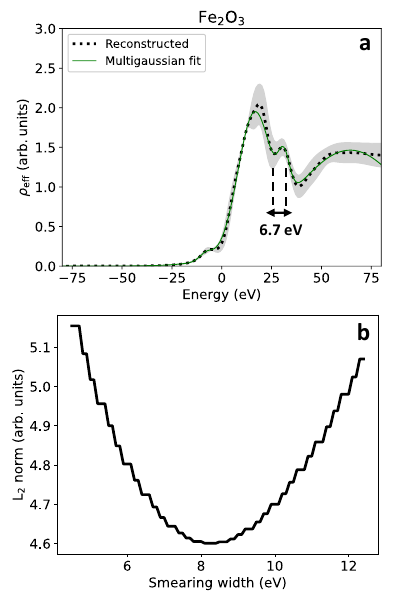}
     \caption{Two different approaches for the estimate of the extracted DOS resolution. In panel (a) this is carried out by fitting a superposition of multiple Gaussians to the reconstructed DOS, whereas in panel (b) we consider the L$_2$ distance between the experimental reconstruction and the DFT simulations as a function of their smearing width.}
     \label{Fig:resolution studies_methods}
 \end{figure}

The experimental reconstructions in Fig.~\ref{Fig:reconstructed} agree well with the theoretical predictions, and we note that we are able to extract densities of states with a fairly complex structure from RIXS spectra with relatively low SNR. We estimated the resolution of the extracted $\rho_{\rm eff}$ with two different methods, shown in Fig.~\ref{Fig:resolution studies_methods}. In panel (a) we fit the reconstructed Fe$_2$O$_3$ $\rho_{\rm eff}$ with a superposition of multiple Gaussians.
This fitting is then used to achieve an estimate of the narrowest feature width we reconstruct, and therefore, of the resolution. This width is found to be 6.7 eV. Another method to evaluate the resolution is to consider the L$_2$ distance between the experimental $\rho_{\rm eff}$ and the DFT predictions, as a function of the smearing width applied to the latter. This curve is plotted in Fig.~\ref{Fig:resolution studies_methods}b. The minimum of this curve can be found for a smearing width around 8.2 eV, yielding another estimate for the resolution of our extracted $\rho_{\rm eff}$. The small discrepancy between the two methods can be traced back to the different resolutions with which different parts of the DOS seem to be reconstructed (e.g. the continuum slope and the small valence band peak).
This resolution is somewhat lower than the limit of 5.5 eV imposed by the spectrometer resolution in our experimental setup.
We attribute this small difference primarily to the SNR of the experimental data, which was fairly low, as the data collection was done in single-shot mode (see Fig.~\ref{Fig:reconstructed}).
Nevertheless, the reconstructed electronic structures show resolutions up to three times higher than the spectral FWHM of the XFEL pulse. We show in Fig.~\ref{Fig:reconstructed}a and b that this allows us to distinguish the environment in which an Fe atom is embedded, via differences in the electronic structure of Fe in the insulating Fe$_2$O$_3$ or in the metallic state. Such a distinction cannot be made via RIXS with the normal spectral resolution of the SASE beam of $\sim$20~eV.


\subsection{Characteristic of the measured electronic structures}

The comparison of the electronic structure with the orbital-projected DOS reveals that the pre-edge feature in Fe$_2$O$_3$, located around -10 eV, contains a mixture of the $2p$ oxygen states and the $3d$ iron states, which are indistinguishable at these resolutions. The band gap between these states and the conduction ones, predicted by experimental measurements~\cite{mohanty1992studies,westre1997multiplet} to have a width of approximately 2 eV, is visible in the reconstruction as a slightly convex plateau located at just below 0 eV. 

In addition to the modulation in the continuum and the pre-edge feature in the Fe$_2$O$_3$ case, we are able to reconstruct the M-shell for both materials, as shown in the insets of Figs.~\ref{Fig:reconstructed}a-b. Despite the small area of these features due to the low thermal de-occupation, they are amplified in the RIXS spectra by their vicinity to the resonant energies. Thus, the reconstructed $\rho_{\rm eff}$ gives us a method to experimentally estimate the binding energy (BE) of the M-shell in extreme thermodynamic conditions—a quantity difficult to obtain via theoretical approaches~\cite{stewart1966lowering, ecker1963lowering}. The values found from the reconstructions are $(-56\pm 3)$~eV for Fe and $(-59\pm 5)$~eV for Fe$_2$O$_3$. As we observed that finite-temperature effects like pressure ionization have a negligible impact on BE for the temperatures reached in the experiment, we compared our estimates with the M-shell binding energies documented in the NIST database~\cite{lemmon2010nist} and we found our results to be compatible with these reference values.

Similarly to what has been done in~\cite{Humphries:2020}, we can exploit the M-shell feature to estimate the sample temperature averaged over the beam pulse duration. However, contrary to~\cite{Humphries:2020}, where the sample temperature is found by fitting the theoretical RIXS spectra to the experimental ones, here we can directly fit the DFT-computed density of accessible states ($\rho|M|^2$) to the reconstructed $\tilde{\rho}_{\rm eff}$:

\begin{equation}
[1-f_{FD}(\epsilon;T)]\rho_{\small{{\rm DFT}}}(\epsilon)|M_{{\rm DFT}}(\epsilon)|^2 \stackrel{!}{=} \tilde{\rho}_{\rm eff}(\epsilon) \ ,
\label{temperature fit}
\end{equation}
where the chemical potential must be computed self-consistently. This new procedure enables us to reduce the error bars on the temperatures estimates compared to the previous work, yielding temperatures of $(6.1\pm 0.2)$~eV for Fe and of $(5.2\pm 0.2)$~eV for Fe$_2$O$_3$. We note that these estimates justify the implicit assumption that depopulation of the M-shell is given mainly by thermal collisions (and not by photoionization), and therefore can be described by the factor $(1-f_{FD})$.






         
    \label{tab:temperature estimates}




\section{Discussion}

We have used a neural surrogate and differentiable programming to establish a machine learning routine that can extract hyper-resolved DOS measurements from RIXS diagnostics at XFEL experiments through large amounts of data. This procedure could represent an alternative way forward in XFEL experiments which prioritizes the increase of the laser pulse energy over the development of monochromation techniques. Our method for the extraction has been compared to other approaches, demonstrating its robustness to noise present in the experimental measurements. We have furthermore shown that the analysis of RIXS spectra through this method can be used to gain a plethora of useful information, from distinguishing between different material spectra to inferring temperatures in HED regimes. A potential future step to further improve the resolution would envisage the inclusion of the setup instrument function in the forward model. This is particularly beneficial in instances where we can model the effects of specific setup components, such as the broadening due to a mosaic crystal~\cite{schlesiger2017new}. Beyond the application to RIXS analysis for further materials and experiments, this machine learning approach can provide a promising avenue for other diagnostics, such as X-Ray Thomson Scattering (XRTS) or the extraction of structure factors from diffraction data. Even though each such application requires a differentiable programming implementation of the corresponding forward model, the combination of machine learning with known physical inductive biases under this scheme constitutes a powerful tool for the analysis of experimental data and the estimation of physical quantities that can only be measured indirectly.


\section{\label{sec:methods} Methods}

The physical parameters of $P_{\rm RIXS}$, chosen for our synthetic and experimental data analysis, are specific to iron and have been taken from~\cite{henke1982low}. We report them in Table~\ref{tab:params}.

\begin{table}[h!]
    \renewcommand{\arraystretch}{1.2}
    \centering
    \scalebox{1}{
    \begin{tabular}{|M{25mm}|M{20mm}|M{20mm}|}
        \hline
        Parameter[units] & $f$ & Value \\\hline\hline
        $\epsilon_K[eV]$ & 1, 2 & -7112\\\hline
        \multirow{2}{*}{$\epsilon_{L,f}[eV]$} & 1 & -723 \\
        & 2 & -708.5 \\\hline
        \multirow{2}{*}{$A_f$} & 1 & 50 \\
        & 2 & 100 \\\hline
        \multirow{2}{*}{$\Gamma_f[eV]$} & 1 & 3 \\
        & 2 & 2.55 \\
        \hline
        
    \end{tabular}}
    \vspace{20pt}
    \caption{Physical parameters used in this work. Iron has two L-shell states, implying two possible final states $f=1,2$.}
    \label{tab:params}
\end{table}

\subsection{\label{sec: ML approach}Details of the machine learning approach}

The estimator used to reconstruct $\rho_{\rm eff}(\varepsilon)$ was a feed-forward neural network with a single input and output, 4 hidden layers with 40 nodes each, and the \textit{softplus} activation function~\cite{zheng2015improving}. The neural network was trained by means of the ADAM~\cite{Kingma2014Adam:Optimization} optimizer with the initial learning rate set to $10^{-3}$. This optimization, together with the automatic differentiation of the forward model, was carried out using the library PyTorch~\cite{ketkar2021introduction}. To ensure a good performance of this scheme, we have identified the necessity of addressing the vanishing gradient problem~\cite{hochreiter1998vanishing} by omitting exponentially small factors in the gradient backpropagation. The training was performed splitting the set of experimental spectra ($I_k$, $\Phi_k$ for $k=0,1,\dots,N$) into batches, whose size was increased dynamically during the training to shift from exploratory to exploitative training. The optimization routine considers the loss function, and the respective gradients, on an entire batch to update model parameters at each step. Furthermore, the training has been carried out over many epochs, i.e. rolling over all the N shots multiple times. Finally, during the training we constrained the output of the neural network $\tilde{\rho}_{\rm eff}(\varepsilon)$ to be positive, a fundamental physical requirement.

\subsection{\label{sec:numerical approach}Details of the numerical approach}

To extract $\rho_{\rm eff}$ from a set of RIXS and SASE spectra without making use of machine learning techniques, we employ the following routine:

\begin{enumerate}
    \item We first discretize the RIXS operator for each FEL spectrum $\Phi_k$, computing the associated matrices $M_k$ for $k=1,2,\dots,N$. Each $M_k$ is constructed taking as its i-th column $P_{\rm RIXS}[\Phi_k,\rho_i]$, where $(\rho_i)_j = \delta_{i,j}$, with $i,j=1,2,\dots,L$ and $L$ the length of $\rho_{\rm eff}$. Note that this discretization is possible because $P_{\rm RIXS}$ is a linear operator in $\rho_{\rm eff}$.
    \item We then construct the macro-linear system of equations in $\rho_{\rm eff}$ by stacking the matrices $M_k$ and the RIXS spectra $I_k$ as:

    \begin{equation}
    \begin{bmatrix}
    M_1 \\
    M_2 \\
    \vdots \\
    M_N
    \end{bmatrix}
    \rho_{\rm eff} \vcentcolon = M \rho_{\rm eff} = 
    \begin{bmatrix}
    I_1 \\
    I_2 \\
    \vdots \\
    I_N
    \end{bmatrix}
    \vcentcolon = I
        \label{linear system}
    \end{equation}

    Notice that the matrix $M$ is in general not square. 
    \item Finally, we approximate the solution of this linear system, and hence $\rho_{\rm eff}$, using the conjugate gradient descent method~\cite{luenberger1984linear}.
\end{enumerate}

This idea of discretizing the physical process operator and inverting the associated matrix has already been employed in the field of spectroscopic analysis~\cite{kayser2019core}. When working with the averaged spectra, we just search for a solution of the linear system $\bar{M}\rho_{\rm eff}=\bar{I}$, where $\bar{M}$ is constructed using the averaged FEL spectrum $\bar{\Phi}$.

\begin{figure*}[t]
     \centering
     \includegraphics[width=\textwidth]{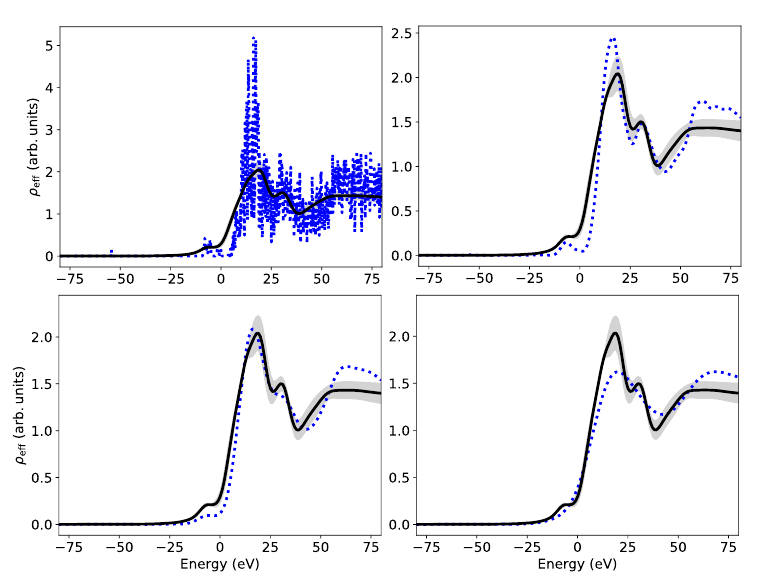}
     \caption{Comparison between the reconstructed $\rho_{\rm eff}$ for Fe$_2$O$_3$ (solid black lines) and the DFT simulations (dashed blue lines) with smearing widths of 0.2, 5, 10, and 20 eV respectively. It is observable that with a smearing of 20 eV, as it would be given by the SASE bandwidth, the characteristic features of Fe$_2$O$_3$ are no longer visible.}
     \label{Fig:resolution studies_comparison}
 \end{figure*}

\subsection{\label{sec: dft calculations}Theoretical calculations}

We simulated the electronic structure of Fe and Fe$_2$O$_3$ using finite-temperature density functional theory (DFT) with a locally-modified version of the \textsc{Abinit} v8.10.3 code~\cite{gonze2016recent,torrent2008implementation,bottin2008large}. The relevant modification is the inclusion of the hybrid Kohn-Sham plane-wave-approximation scheme~\cite{zhang2016extended} as described by Ref.~\cite{blanchet2022density} to enable accurate high temperature calculations.
The \textsc{Abinit} code is used to solve the Kohn-Sham (KS) system~\cite{KohnSham}, which provides the KS states and eigenvalues. The DOS is given by the latter, while the former can be used to calculate the dipole transition matrix elements from the Fe $1s$ core state to the calculated valence states~\cite{mazevet2010calculations}. The orbital-projected DOS was also calculated to aid in identifying features in the total DOS.

The ion cores were represented using the projector-augmented wave (PAW) scheme~\cite{BlochlPAW}, with Fe and O PAW potentials generated using the \textit{Atompaw} code~\cite{atompaw}. For efficiency, the Fe $1s$, $2s$, $2p$ orbitals and the O $1s$ orbital were treated with the frozen-core approximation. This approximation is suitable for the relatively low temperatures ($< 10$~eV) reached in this experiment, as these orbitals are not thermally ionized.

Calculations were performed at electron temperatures of 300~K, 1~eV, 5~eV, and 10~eV, with the ions frozen in their ambient crystal lattice positions. This further approximation is justified by the femtosecond duration of the XFEL pulses, during which all the spectral emission of interest occurs. This timescale is substantially shorter than the electron-phonon coupling times of several picoseconds.
The Fe calculations were therefore performed in a bcc primitive unit cell containing a single atom. Simulations were carried out with 120 bands, a $30 \times 30\times 30$ $k$-point grid, with a cut-off energy of 50~Ha for the PAW pseudo-wavefunctions and 150~Ha for the all-electron wavefunctions.
The Fe$_2$O$_3$ calculations were performed in its $\alpha$-phase~\cite{sakurai2009first}, with a unit cell containing 30 atoms, 1440 bands, a $6 \times 6 \times 1$ $k$-point grid (the latter direction being the long direction of the unit cell), with a cut-off energy of 20~Ha for the PAW pseudo-wavefunctions and 100~Ha for the all-electron wavefunctions.

For the exchange-correlation functional, the PBE form of the generalized gradient approximation (GGA) was used for both the Fe and O atoms. 
We used the corrective approach of a Hubbard potential (DFT+$U$) to recover the band gap in Fe$_2$O$_3$, with a value of $U=4.0$~eV for the Fe $l=2$ channel to be able to recover a band gap of $\sim 2.0$~eV, as given by~\cite{xia2013tuning}. 
As the higher temperatures we consider here are substantial compared with the value of $U$, we choose $U=4.0$~eV for all Fe$_2$O$_3$ calculations. Small adjustments to the value of $U$ at a temperature of 5~eV did not show meaningful changes to the electronic structure, justifying this approach.

\section*{Data availability}
The data that support the findings of this study are available from the
corresponding author upon request.

\section*{Code availability}
The code for the simulation of the self-referenced arrival time spectra
is available from the corresponding author upon request.

\section*{Acknowledgements}
A.F. and S.M.V. acknowledge support for the STFC UK Hub for the Physical Sciences on XFELS.
T.G., P.S., J.S.W. and S.M.V. acknowledge support from AWE via the Oxford Centre for High Energy Density Science (OxCHEDS).
C.C., S.A., J.S.W. and S.M.V. acknowledge support from the UK EPSRC under grants EP/P015794/1 and EP/W010097/1.
K.K.A.E, T. C. and S.M.V. acknowledge support from the Royal Society. The work of D.N.P., D.A.C. and E.S. was supported by the Department of Energy [National Nuclear Security Administration] University of Rochester “National Inertial Confinement Fusion Program” under Award Number(s) DE-NA0004144.

\section*{Contributions}
S.M.V., J.S.W., A.F., M.H., G.W.C., D.M., O.S.H., T.R.P, C.B. and T.G. conceived and designed the experiment. A.F., S.M.V., T.G., O.S.H., T.R.P., C.B., C.C., T.C., P.S., S.A., P.H., Y.S, D.A.C., E.S and D.N.P. carried out the experiment and the real-time data analysis. V.B. and H.H. contributed to the beamline control. A.F, S.M.V. and K.K.A developed the deconvolution approach. A.F. conducted the studies on the synthetic data with contributions from K.K.A. A.F. performed the post-experiment data analysis with help from T.G., D.N.P and S.A. T.G. performed the DFT simulations. A.F., T.G, K.K.A and S.M.V wrote the manuscript with contributions from all authors.

\section*{Competing interests}
The authors declare no competing interests.

\end{document}